\documentclass[aps,twocolumn,showpacs]{revtex4}%
\usepackage{amsfonts}
\usepackage{amsmath}
\usepackage{amssymb}
\usepackage{graphicx}
\usepackage{color}%
\providecommand{\U}[1]{\protect\rule{.1in}{.1in}}
\begin{document}
\title{Controllable unidirectional transport and photon storage in an one-dimensional non-Hermitian lattice with complex hopping rates}

\author{Lei Du$^{\dag}$ and Yan Zhang}

\affiliation{Center for Quantum Sciences and School of Physics, Northeast Normal University, Changchun 130117, P. R. China}

\affiliation{$^{\dag}$leiduphys@outlook.com}

\date{\today }


\begin{abstract}
We study an one-dimensional non-Hermitian lattice with complex hopping rates, which can be realized by a quasi-one-dimensional
sawtooth-type Hermitian lattice after adiabatic elimination. By means of synthetic magnetic fluxes, the imaginary parts of the complex
hopping rates can be modulated by additional phase, thus a non-reciprocal structure arises. With such a non-Hermitian lattice, one can
realize robust unidirectional transport for both single-site and Gaussian excitations, which is immune to defects and backscattering.
Furthermore, we proposed a sandwich structure based on the non-Hermitian lattice, which can be used for realizing controllable photon
storage and reversal. The storage time and range can be artificially controlled within limits, and the storage efficiency can be increased
via a finite gain compensation. The proposal of the non-Hermitian lattice in this paper opens up a new path for controllable unidirectional
photon transport and provides a promising platform for quantum information process (QIP).
\end{abstract}
\maketitle

\section{Introduction}

The realization of controllable photon transport is vital for a large variety of areas and has been a major goal of research for decades \cite{angular-momen}.
Especially, unidirectional photon transport, which can be used for realizing optical isolators and diodes, plays a key role in modern optics. Generally,
unidirectional wave transport can be observed in an asymmetric hybrid system in which a nonlinear subsystem is embedded in a linear medium \cite{non-oneway1,
non-oneway2,asym}, or a two- (three-) dimensional photonics lattice with topological protection \cite{topo1,topo2,topo3}. As is well-known, topologically protected
edge states, which are guided by synthetic gauge fields and propagate at the boundary of the systems, have exhibited enormous advantages due to the robustness,
i.e., they are immune to disorders and defects. However, such schemes are realized only in two- or three-dimensional photonics lattices such as quantum Hall
systems, topological insulators and topological superconductors, rather than one-dimensional photonics lattice.

An alternative platform for realizing controllable and unidirectional photon transport is the non-Hermitian photonic lattice, in which a large amount
of important phenomena have been observed, such as non-Hermitian delocalization in disorder lattices \cite{H-N,delo1,delo2,delo3,delo4,delo5,delo6,
delo7}, invisible defects \cite{invisdefect}, topological phase transitions \cite{topophase1,topophase2,topophase3}, non-Hermitian induced flat bands
\cite{flat1,flat2,flat3}, invisible non-Hermitian optical potentials \cite{KKpra,KKol} and so on. Recently, inspired by the non-Hermitian delocalization
proposed by Hatano and Nelson \cite{H-N}, schemes of realizing robust unidirectional photon transport in one-dimensional non-Hermitian lattice was
proposed by Longhi \emph{et al.} \cite{longhiSR,oneway}.  In those works, an imaginary gauge field, which can be acquired by exploiting auxiliary ring
resonators with gain and loss in different half perimeters respectively, leads to an non-reciprocal hopping rates. By means of the imaginary gauge field,
the wave is amplified in one direction, whereas it is attenuated in the opposite direction, namely, unidirectional photon transport is realized. Subsequently,
a non-Hermitian lattice with imaginary hopping rates, which can be used for realizing robust bidirectional photon transport, is proposed in Refs.~\cite{engineer,
longhi-bi}. These inspiring works enable non-Hermitian lattices to be feasible and excellent platforms for realizing controllable and robust photon transport.

In this paper, we show that a class of one-dimensional non-Hermitian lattices, with complex hopping rates whose imaginary parts are modulated by synthetic magnetic
fluxes, support robust unidirectional photon transport which is reflectionless and is immune to defects. Such a non-Hermitian lattice can be realized by a tailored
quasi-one-dimensional sawtooth-type Hermitian lattice after adiabatic elimination. By adjusting the synthetic magnetic fluxes, a filter with selectable wave numbers
and a switch between localization and delocalization can be realized. Furthermore, we also show a sandwich structure which is based on the non-Hermitian proposal,
can be utilized for a photon storer. By modulating relevant parameters, one can realize both photon storage and reversal. The storage time and range can be artificially
controlled within limits, and the storage efficiency can be increased by several schemes with finite gain compensation.

\section{Model and equations}

\begin{figure}[ptb]
\centering
\includegraphics[width=8.5 cm]{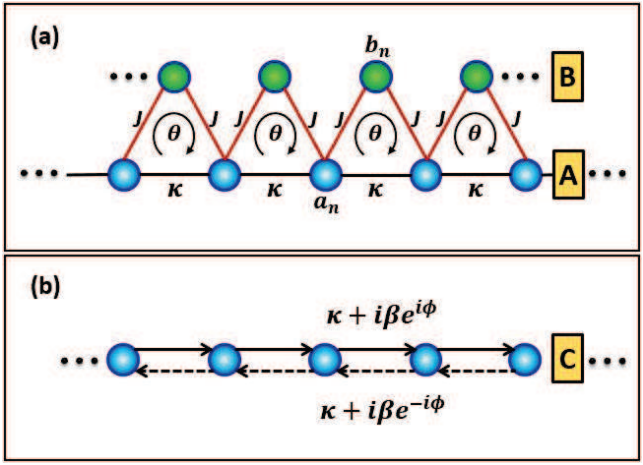}
\caption{(Color online) Schematic illustration of (a) the quasi-one-dimensional sawtooth-type lattice which serves as the physical
realization of the non-Hermitian lattice proposed in this paper and (b) the equivalent one-dimensional non-Hermitian Lattice with
complex hopping rates whose imaginary parts are modulated by synthetic magnetic fluxes. Here, $\kappa$ is the hopping rate between adjacent
sites on the sublattice $A$ and $t$ is the hopping rate between the nearest-neighbor sites on the sublattices $A$ and $B$, respectively.}\label{fig1}
\end{figure}

We consider photon transport in a quasi-one-dimensional sawtooth-type lattice, consisting of a main sublattice $A$ with optical
potential $U_{a,n}$ and Wannier state $|n\rangle_{A}$, and an auxiliary sublattice $B$ with optical potential $U_{b,n}$ and Wannier
state $|n\rangle_{B}$, at the $n$th sites of $A$ and $B$ respectively, as schematically shown in Fig.~\ref{fig1} (a). As in general optical
lattices, the real parts of potentials $V_{a (b),n}=$Re$[U_{a (b),n}]$ denote detunings (coupled-resonator systems) or propagation constants
offsets (optical-waveguide systems), and the imaginary parts $\gamma_{a (b),n}=$Im$[U_{a (b),n}]$ denote gain or loss effects. Hereafter,
we assume an uniform loss rate $\Gamma$, i.e., $U_{a,n}=V_{a,n}-i\Gamma$, for the whole sublattice $A$. Indicating by $a_{n}$ and $b_{n}$
the probability amplitudes of the $n$th sites of sublattices $A$ and $B$, the evolution equations of the system in the tight-binding approximation
can be written as
\begin{eqnarray}
i\frac{d a_{n}}{d t}&=&U_{a,n}a_{n}+\kappa(a_{n+1}+a_{n-1})+J(e^{i\theta}b_{n}+e^{-i\theta}b_{n-1})\nonumber\\
i\frac{d b_{n}}{d t}&=&U_{b,n}b_{n}+J(e^{i\theta}a_{n+1}+e^{-i\theta}a_{n})
\label{eq1}
\end{eqnarray}
where $\kappa$ is the hopping rate between the nearest-neighbor sites of sublattice $A$ and $J$ is the nearest-neighbor hopping rate
between sites of sublattices $A$ and $B$, respectively. We set $\theta$ the synthetic magnetic fluxes perpendicular to the plane of
the quasi-one-dimensional lattice, which can be acquired by exploiting synthetic gauge fields \cite{longhi-bi,imbalance,optomag}. A simple
and feasible scheme for generating synthetic gauge fields by means of optomechanics on a lattice is proposed in Ref.~\cite{optomag}. Moreover,
synthetic gauge fields can also be introduced by optical path imbalances between adjacent sites \cite{imbalance} or modulated on-site potentials
\cite{ABcage}. By assuming an uniform optical potential $U_{b}$, the auxiliary sublattice $B$ can be eliminated adiabatically, i.e.,
\begin{eqnarray}
b_{n}\simeq \frac{-J[e^{i\theta}a_{n+1}+e^{-i\theta}a_{n}]}{U_{b}}
\label{eq2}
\end{eqnarray}
if $|U_{b}|$ is much larger than $J$ and energy $E$ (in modulus) of the system ($-2\kappa\leq |E|\leq 2\kappa$) \cite{longhi-bi,engineer}.
Then the equivalent evolution equations can be read
\begin{eqnarray}
i\frac{d a_{n}}{d t}=U_{eff,n}a_{n}+J_{1}a_{n+1}+J_{2}a_{n-1}
\label{eq3}
\end{eqnarray}
where $U_{eff,n}=U_{a,n}-2J^{2}/U_{b}$ denotes the effective optical potential at the $n$th site of $A$. $J_{1}=\kappa-J^{2}e^{2i\theta}/U_{b}$
and $J_{2}=\kappa-J^{2}e^{-2i\theta}/U_{b}$ denote the non-reciprocal effective hopping rates, which break the time-reversal symmetry of the
sublattice $A$. Note, equivalent complex hopping rates $\kappa+i\beta e^{\pm2i\theta}$, which the imaginary parts are modulated by synthetic magnetic
fluxes, are introduced if $U_{b}=iJ^{2}/\beta$. It is so different from the general Hermitian lattices where the hopping rates are real and this will
be discussed in detail in the next section. In view of this, we can obtain an equivalent one-dimensional non-Hermitian lattice $C$ with complex hopping
rates, as shown in Fig.~\ref{fig1} (b). The evolution equations of the non-Hermitian lattice $C$ can be written as
\begin{eqnarray}
i\frac{d c_{n}}{d t}=-i\gamma c_{n}+(\kappa+i\beta e^{i\phi})c_{n+1}+(\kappa+i\beta e^{-i\phi})c_{n-1}
\label{eq4}
\end{eqnarray}
with $\phi=2\theta$ and $\gamma=\Gamma-2\beta$. For the sake of convenience, we have removed the real parts of $U_{eff,n}$ by setting $V_{a,n}=0$, because a
nonvanishing real part only shifts the dispersion relation. Such a set of coupled evolution equations describe the dynamics of the non-Hermitian lattice with
complex hopping rates whose imaginary parts are modulated by an additional Peierls phase $\phi$ (the synthetic magnetic fluxes). The phase $\phi$ can be
precisely adjusted for several implementation schemes mentioned above, it seems that a phase-dependent control of light propagation can be realized.

\begin{figure}[ptb]
\centering
\includegraphics[width=8.5 cm]{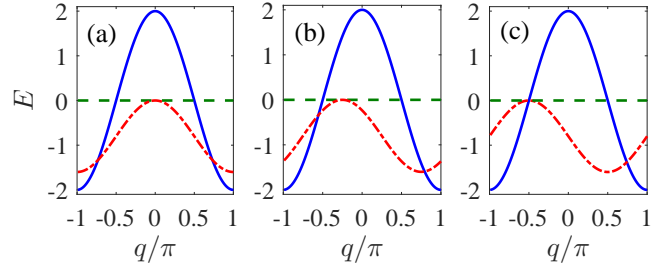}
\caption{(Color online) The real part (blue solid line) and the imaginary part (red dot-dashed line) of
the energy dispersion curve versus the Bloch wave number $q$ with (a) $\phi=0$; (b) $\phi=\pi/4$; (c) $\phi=\pi/2$.
The green dashed line denotes $E=0$ as a reference line. Other parameters are $\gamma=0.8\kappa$ and
$\beta=0.4\kappa$. The energy scale is in arbitrary units.}\label{fig2}
\end{figure}

By setting the solutions of Eq.~(\ref{eq3}) in form $c_{n}=C$exp$(iqn-iEt)$, the dispersion relation of the non-Hermitian lattice can be written as
\begin{eqnarray}
E(q)=2\kappa \cos{(q)}+2i\beta \cos{(q+\phi)}-i\gamma
\label{eq5}
\end{eqnarray}
with $-\pi\leq q\leq\pi$ is the Bloch wave number (quasi-momentum) in the first Brillouin zone. One can find that the imaginary part of the dispersion
relation which describes the amplification or absorption effect, can be modified by the phase $\phi$, while the real part which determines the
group velocity, i.e.,
\begin{eqnarray}
v_{g}=\text{Re}(\frac{d E}{d q})=-2\kappa\sin{(q)}
\label{eq6}
\end{eqnarray}
is invariable versus $\phi$. Note that a purely dissipative optical system requires $\gamma\geq2\beta$ \cite{longhi-bi}. Moreover, we point out that a
Hermitian lattice arises when $\gamma=\beta=0$.

We show in Fig.~\ref{fig2} the dispersion curve as the function of $q$. As predicted in Eq.~(\ref{eq4}), the position of imaginary part of the
dispersion relation (especially the maximal value Im$(E)_{max}=0$ which corresponds to a lossless light propagation) can be changed by adjusting the
phase $\phi$ while the real part keeps invariable. It seems somewhat similar to the case in Ref.~\cite{longhi-bi} that the relative position can be
adjusted by the phase $\phi$. However, due to the movable imaginary part of the dispersion relation, a filter with selectable wave numbers can be realized,
as shown in the next section.

\begin{figure}[ptb]
\centering
\includegraphics[width=8.8 cm]{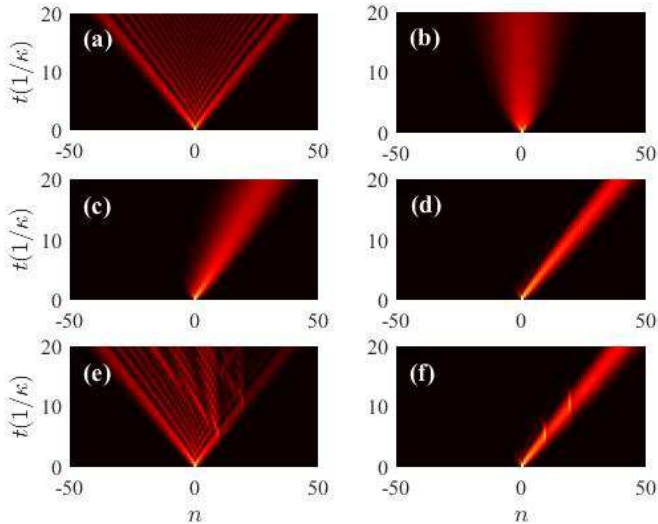}
\caption{(Color online) Dynamics of wave propagation in Hermitian and non-Hermitian lattices with single-site excitation at initial time. We plot
in figures the space-time evolution of the normalized probability amplitude $|\rho_{n}(t)|$ in Hermitian lattices without [(a)] and with [(e)] defects,
non-Hermitian lattices without [(b)-(d)] and with [(f)] defects. We set $\gamma=\beta=0$ in the Hermitian case and $\beta=0.4\kappa, \gamma=0.8\kappa$
in the non-Hermitian case. The lattice defects are assumed to be $V_{c}=2\kappa$ at $n=10,20$. Other parameters are (b) $\phi=0$, (c) $\phi=\pi/4$, (d)
and (f) $\phi=\pi/2$. In the Hermitian case the dynamics do not depend on the phase $\phi$.}\label{fig3}
\end{figure}

\section{Robust unidirectional photon transport}

We first consider wave propagation in the non-Hermitian lattice $C$ with single-site excitation at initial time $t=0$, i.e., $c_{n}(0)=\delta_{n,0}$.
Fig.~\ref{fig3} shows numerically the space-time evolution of the normalized probability amplitude $\rho_{n}(t)$,
\begin{eqnarray}
\rho_{n}(t)=\sqrt{\frac{|c_{n}(t)|^{2}}{S(t)}}, \,\,S(t)=\sum_{n}|c_{n}(t)|^{2}
\label{eq7}
\end{eqnarray}
\begin{figure}[ptb]
\centering
\includegraphics[width=8.5 cm]{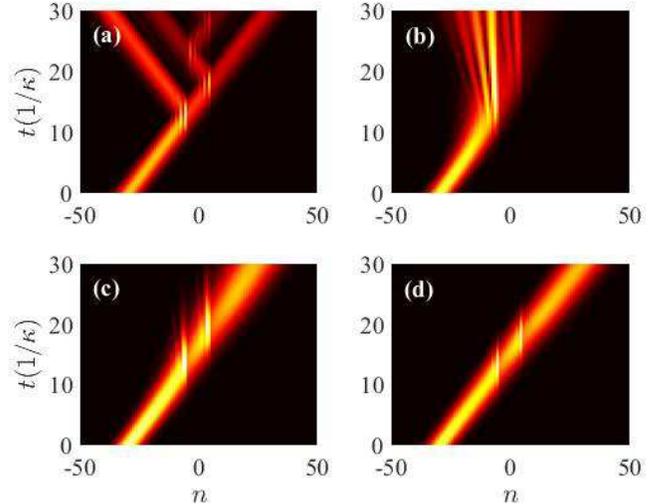}
\caption{(Color online) Dynamics of wave propagation in Hermitian and non-Hermitian lattices with Gaussian excitation at initial time. We plot
in figures the space-time evolution of the normalized probability amplitude $|\rho_{n}(t)|$ in Hermitian lattices [(a)] and non-Hermitian lattices
[(b)-(d)] with defects. We set $\gamma=\beta=0$ in the Hermitian case and $\beta=0.4\kappa, \gamma=0.8\kappa$ in the non-Hermitian case. The
lattice defects are assumed to be $V_{c}=2\kappa$ at $n=\pm5$. We assume here a Gaussian wave packet with $n_{0}=-30, w_{0}=5, q_{0}=-\pi/2$.
Other parameters are (b) $\phi=0$, (c) $\phi=\pi/4$ and (d) $\phi=\pi/2$. In the Hermitian case the dynamics do not depend on the phase $\phi$.}\label{fig4}
\end{figure}
Note a single-site excitation refers to an incident wave including components with all Bloch wave numbers in the first Brillouin zone. So for
a general Hermitian lattice ($\gamma=\beta=0$), the incident wave propagates with all velocities in the range $[-2\kappa, 2\kappa]$, as shown in
Fig.~\ref{fig3} (a), an expanding sector-type propagation pattern arises instead of a beam-type pattern. However, for a non-Hermitian lattice
with complex hopping rates, owing to the existence of the imaginary dispersion relation, the incident wave propagates only with the velocity corresponding
to Im$(E)=0$. As shown in Fig.~\ref{fig3} (b)-(d), by adjusting the phase $\phi$, one can realize a transition between localization and delocalization
of incident wave and a control of group velocity. Physically, this is because the group velocity corresponding to Im$(E)=0$ is vanishing when $\phi=0$,
incident components with other wave numbers decay rapidly during propagation. By adjusting the phase $\phi$, the lossless incident component acquire
a nonvanishing velocity. The group velocity of the right-propagating wave in Fig.~\ref{fig3} (b)-(d) tends to the maxima $v_{g}=2\kappa$ as $\phi\rightarrow\pi/2$.
In addition, a left-propagating wave with similar behavior can also be observed by adjusting $\phi$ to be negative, according to Eqs.~(\ref{eq5}) and (\ref{eq6}).

\begin{figure}[ptb]
\centering
\includegraphics[width=8.5 cm]{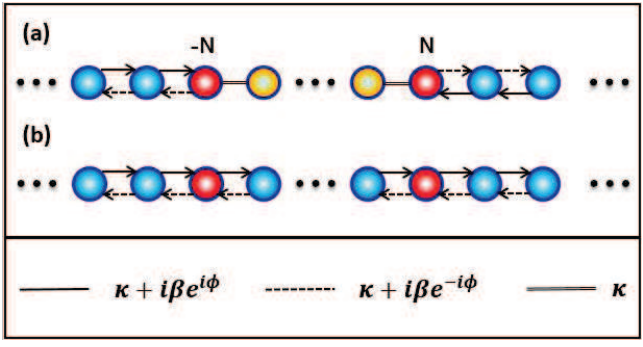}
\caption{(Color online) Schematic illustration of the photon storage process: (a) The three-stage heterogeneous structure used for photon capture and (b)
defective homogeneous non-Hermitian lattice used for photon release. Blue and yellow circles indicate sites of the non-Hermitian and the Hermitian parts,
respectively. Red circles indicate the boundary sites with defect $V_{c}$. The hopping rates of the two semi-infinite non-Hermitian parts in (a) are asymmetric
due to the opposite synthetic magnetic fluxes, whereas they are symmetric in (b). The hopping rates of the Hermitian part in (a) are real and identical.}\label{fig5}
\end{figure}

Robust unidirectional photon transport can be observed in the non-Hermitian lattice by introducing lattice defects, i.e., additional potential terms
such as $V_{add}=V_{c}(\delta_{n,10}+\delta_{n,20})$ with $V_{c}=2\kappa$ in Eq.~(\ref{eq4}), as shown in Fig.~\ref{fig3} (e) and (f). One can find in Fig.~\ref{fig3} (e)
that the right-propagating wave in the Hermitian lattice undergos multiple transmissions and reflections at the two defects like in a Fabry-Per\'{o}t cavity, whereas
in Fig.~\ref{fig3} (f) the wave propagation is much more robust in the non-Hermitian lattice, immune to the defects. The physical reason thereof is as follows: In Hermitian
lattices, the dispersion relation is the same as the real part of the non-Hermitian case in Fig.~\ref{fig2} but with a vanishing imaginary part. The energy band is degenerate
and there always exists a reflected wave with the wave number $q'$ satisfying $q'=-q_{0}$ and $E(q')=E(q_{0})$, where $q_{0}$ is the incident wave number. It means that a
reflected wave resulting from elastic scattering at the defects is allowed to propagate in this case. However, in the non-Hermitian case, due to the existence of the imaginary
dispersion relation, the energy band becomes no longer degenerate as long as $\phi\neq0$. The reflected waves become \emph{evanescent} because there is no real solution for the
equation $E(q')=E(q_{0})$ except for $q'=q_{0}$ \cite{oneway,longhi-bi}. Therefore a robust unidirectional photon transport which is immune to defects can be observed in the
non-Hermitian lattice.

The robustness of photon transport can also hold for the non-Hermitian lattice with a Gaussian excitation
\begin{eqnarray}
c_{n}(0)\propto \text{exp}[-\frac{(n-n_{0})^{2}}{w_{0}^{2}}+iq_{0}n]
\label{eq8}
\end{eqnarray}
where $n_{0}$ and $q_{0}$ are respectively the incident lattice site and the central Bloch wave number of the incident wave, and $w_{0}$ is the width of the
Gaussian wave packet. Differing from single-site excitation, Gaussian wave packet contains only a narrow range of Bloch wave number. Similar to Fig.~\ref{fig3} (e),
one can find in Fig.~\ref{fig4} (a) that the right-propagating wave in the Hermitian lattice is multiply scattered by the defects and the scattering waves decay rapidly.
As a comparison, the behavior in the non-Hermitian lattice is shown in Fig.~\ref{fig4} (b)-(d), where the photon transport tends to be more and more robust as the phase
$\phi$ varies from $0$ to $\pi/2$. Similar behaviors can also be observed for a left-propagating Gaussian wave packet but with $\phi\rightarrow -\phi$ in Eq.~(\ref{eq4}).

\begin{figure}[ptb]
\centering
\includegraphics[width=8.5 cm]{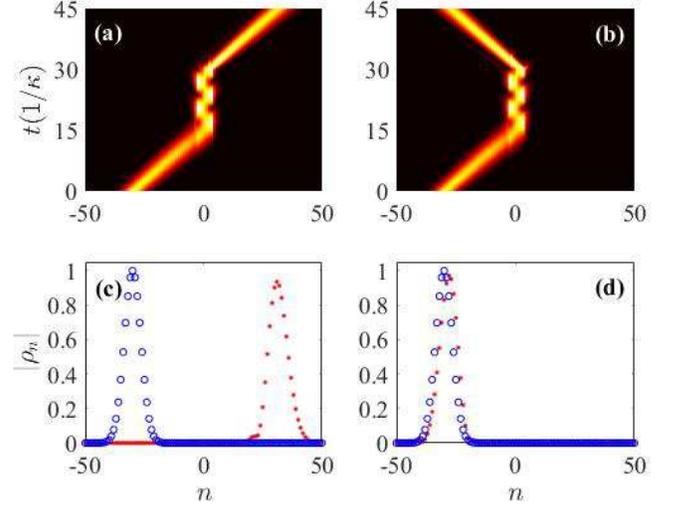}
\caption{(Color online) Storage and reversal of a left incident Gaussian wave packet in the three-stage heterogeneous structure. (a) Space-time evolution of photon
storage process. (b) Space-time evolution of photon reversal process. (c) and (d) Normalized amplitude profile $|\rho_{n}|$ versus the lattice sites $n$ for storage
and reversal processes, respectively. The Gaussian excitation we used here is the same as the one in Fig.~\ref{fig4}. The switch time in (a) and (b) is $t'\kappa=30$.
The defects at the boundary sites $n=\pm N$ are $V_{c}=\kappa$ where $N=3$. Other parameters in Eqs.~(\ref{eq9}) and (\ref{eq10}) are the same as those in
Fig.~\ref{fig4}.}\label{fig6}
\end{figure}

Physically, for the non-Hermitian lattice with $\phi=0$, the dispersion curve is symmetric with respect to $q=0$ in the first Brillouin zone, similar to that
in the Hermitian case, but with a nonvanishing imaginary part. Both transmitted and reflected waves can be observed like in the defective Hermitian lattice. The scattered
waves decay rapidly because the central incident wave number $q_{0}=-\pi/2$ in Fig.~\ref{fig4} (b) corresponds to an attenuate propagation in this case. As $\phi\rightarrow\pi/2$,
according to Eq.~(\ref{eq5}), the imaginary part of the dispersion relation tends to be vanishing for $q_{0}=-\pi/2$. The energy band losses the degeneracy and thereby the
reflected waves become \emph{evanescent}, i.e., the wave numbers of the reflected waves are complex \cite{oneway,longhi-bi}. In view of this, robust unidirectional photon
transport which is immune to defects can also be realized in the non-Hermitian lattice for Gaussian excitations.

It is worth to note that Gaussian wave packets with other central wave numbers can also propagate robustly in the non-Hermitian lattice by adjusting the phase
to satisfy $\phi=-q_{0}$, according to Eq.~(\ref{eq5}), i.e., a filter with selectable wave numbers can be realized. It is very different from Hermitian lattices
and other non-Hermitian lattices without phase-modulated imaginary hopping rates. Furthermore, due to the analogy between defective lattices and arrays of Fabry-Per\'{o}t
cavities. The robust photon transport scheme can be extended to cavity quantum electrodynamics (CQED) realm.

\section{Photon storage and reversal}

Besides robust unidirectional transport, the non-Hermitian lattice can also be utilized as the building block of a photon storer. To achieve photon storage, we consider a
system which can be switched between a sandwich (i.e., three-stage heterogeneous) structure and a defective homogeneous non-Hermitian lattice, as shown in Fig.~\ref{fig5}.
The sandwich structure consists of two semi-infinite non-Hermitian lattices and a finite Hermitian lattice embedded between them, as shown in Fig.~\ref{fig5} (a). The phase
(synthetic magnetic fluxes) of the two non-Hermitian lattices which modulate the imaginary parts of the hopping rates are opposite, leading to an asymmetric structure.
There exist two identical defects $V_{c}$ at the boundary sites of $n=\pm N$ between the Hermitian and the non-Hermitian parts, respectively. The Hermitian part can be
realized by exploiting additional auxiliary sites for each site of a non-Hermitian lattice in this area. By adjusting the potentials of the additional auxiliary
sites, one can acquire opposite imaginary hopping rates which can offset the original imaginary parts. We excite such a sandwich structure at initial time $t=0$ by a
Gaussian wave packet with central Bloch wave number $q_{0}$. For a left incident wave, we set the phase $\phi=-q_{0}$. Taking into account all the conditions above, the
evolution equations of the system can be written as
\begin{widetext}
\begin{eqnarray}
i\frac{d c_{n}}{d t}=\left\{
\begin{aligned}
&-i\gamma c_{n}+(\kappa+i\beta e^{-iq_{0}})c_{n+1}+(\kappa+i\beta e^{iq_{0}})c_{n-1}&&&n<-N\\
&(V_{c}+i\xi)c_{n}+\kappa c_{n+1}+(\kappa+i\beta e^{iq_{0}})c_{n-1}&&&n=-N\\
&\kappa(c_{n+1}+c_{n-1})&&&-N<n<N\\
&(V_{c}+i\xi)c_{n}+(\kappa+i\beta e^{iq_{0}})c_{n+1}+\kappa c_{n-1}&&&n=N\\
&-i\gamma c_{n}+(\kappa+i\beta e^{iq_{0}})c_{n+1}+(\kappa+i\beta e^{-iq_{0}})c_{n-1}&&&n>N
\label{eq9}
\end{aligned}
\right.
\end{eqnarray}
\end{widetext}
Note the sudden change of the dispersion relation at the boundary sites gives rise to an amplification or attenuation of the wave. According to Eq.~(\ref{eq9}), whenever the
wave propagated in the Hermitian area is reflected by the defects, the imaginary parts of the dispersion relations of the boundary sites, i.e., Im$[E_{n=N}(q_{0})]=$Im$[E_{n=-N}(-q_{0})]=i\beta \cos{(2q_{0})}$,
lead to an amplification or attenuation which depends on the value of $q_{0}$. The effect of the process is non-negligible due to the multiple reflections in the Hermitian area, as
will be discussed below. In view of this, we introduce additional imaginary potential $i\xi$ for the boundary sites to offset the amplification or attenuation.

We set $\xi=\beta$ in Fig.~\ref{fig6} due to the central wave number of the Gaussian wave packet we used here is $q_{0}=-\pi/2$. As shown in Fig.~\ref{fig6} (a), the left incident Gaussian
wave packet propagates unidirectionally in the first non-Hermitian part, as discussed in the previous section. Once the incident wave enters into the Hermitian part, it is scattered
back and forth between the two defects. The photons are thus captured in the Hermitian area. By adjusting potentials of the additional auxiliary sites and the phase $\phi$ (to be
identical over the whole lattice) at $t=t'$ [$(t'\kappa=30)$ in Fig.~\ref{fig6}], the system becomes a defective homogeneous non-Hermitian lattice as shown in Fig.~\ref{fig5} (b), i.e.,
\begin{eqnarray}
i\frac{d c_{n}}{d t}&=&-i\gamma c_{n}+(\kappa+i\beta e^{i\phi})c_{n+1}+(\kappa+i\beta e^{-i\phi})c_{n-1}\nonumber\\
&&+V_{c}(\delta_{n,-N}+\delta_{n,N})
\label{eq10}
\end{eqnarray}
then one can find a restoring wave propagate in the same direction of the incident wave, with $\phi=-q_{0}$. We plot in Fig.~\ref{fig6} (c) the normalized amplitude profiles of the incoming
and outcoming waves. The outcoming wave keeps the Gaussian shape meaning the storage in this system is almost shape-preserving. Photon reversal can also be observed if the phase of the defective
homogeneous non-Hermitian lattice satisfies $\phi=q_{0}$ in Eq.~(\ref{eq10}), as shown in Fig.~\ref{fig6} (b). The wave propagates in the opposite direction after retrieval. We plot
in Fig.~\ref{fig6} (d) the normalized amplitude profiles of the incoming and outcoming waves in this case and once again we find an almost shape-preserving outcoming wave. Similar behaviors
can also arise for a right incident wave but with $q_{0}\rightarrow -q_{0}$ in Eq.~(\ref{eq9}) and $\phi\rightarrow -\phi$ in Eq.~(\ref{eq10}). Note that the storage time and the storage
range (width of the Hermitian area) can be artificially controlled by adjusting the switch time $t'$ and the boundary position $\pm N$ within limits, respectively. A similar scheme, i.e.,
realizing photon storage in a sandwich structure, has been proposed in an atomic system \cite{jhwu2009}. Where the incident wave undergos similar multiple scattering back and forth
in the middle area of the system. In view of this, photon storage and reversal with controllable storage time and range can be realized in the non-Hermitian lattice.

The physical mechanism underlying is that, once the incident wave enters into the Hermitian part of the sandwich structure, it undergos multiple reflections between the defects because
reflections are allowed to propagate in the Hermitian lattice as discussed above. However, transmissions are prevented because the energy bands are asymmetry with respect to the boundary
sites. There is no real solution for the equation $E(q_{0})=E'(q_{0})$, where $E$ and $E'$ are dispersion relations (quasi-energies) of different parts, i.e., transmissions are \emph{evanescent}.
Therefore the wave is reflected back and forth, localized in the Hermitian area. Once the sandwich structure is switched to the defective homogeneous non-Hermitian lattice, due to the robustness
of photon transport in such a lattice as discussed above, the defects can no longer prevent transmissions and thereby the wave is retrieved. Similar analysis also holds for a right incident wave.

\begin{figure}[ptb]
\centering
\includegraphics[width=8.5 cm]{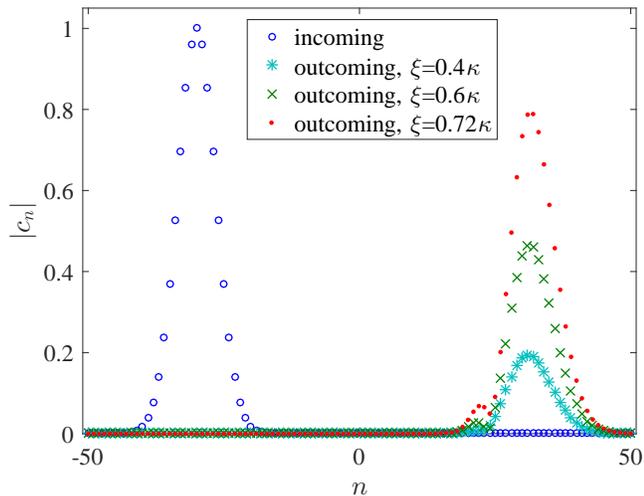}
\caption{(Color online) Profiles of the actual probability amplitude $|c_{n}|$ of the incoming and outcoming waves with different offset coefficient $\xi$ of the boundary sites.
All the parameters except for $\xi$ are the same as those in Fig.~\ref{fig6}.}\label{fig7}
\end{figure}

In order to show the storage efficiency of the photon storage scheme, we plot in Fig.~\ref{fig7} the actual amplitude profiles of the incoming and outcoming waves of the case in Fig.~\ref{fig6}.
One can find that the actual amplitude of the outcoming wave with $\xi=\beta=0.4\kappa$ is much weaker than the incoming one, although we have introduced additional gain for the boundary sites.
This is because the propagation of a Gaussian wave packet like Eq.~(\ref{eq8}) in the non-Hermitian lattice is not exactly lossless due to the width $w_{0}$. In spite of this, as shown in Fig.~\ref{fig7},
one can increase the storage efficiency by enhancing the offset coefficient $\xi$ of the boundary sites. With a increasing $\xi$, the actual amplitudes of the outcoming waves enhance significantly,
and the shape-preserving effect is satisfactory. Moreover, one can also increase the storage efficiency by applying additional gain for the finite sites in the Hermitian area. That is to say, the
photon storage scheme can be optimized by means of finite gain.

\section{Conclusions}

In summary, we study a non-Hermitian lattice with complex hopping rates whose imaginary parts are modulated by synthetic magnetic fluxes. Such a non-Hermitian lattice can be realized
by a quasi-one-dimensional sawtooth-type Hermitian lattice, consisting of a main sublattice and an auxiliary sublattice. With appropriate conditions, the auxiliary sublattice can be
eliminated adiabatically and thereby the complex hopping rates are introduced effectively. With this lattice, robust and reflectionless unidirectional photon transport can be realized,
in spite of the defects on the lattice. By adjusting the phase (synthetic magnetic flux), one can realize an optical filter with selectable wave numbers and a switch between localization
and delocalization. Furthermore, the non-Hermitian lattice can also be used as the building block of a photon storer. By modulating relevant parameters especially the phase, photon storage
and reversal can be observed. The storage time and range can be artificially controlled within limits which are rather difficult in other schemes. The storage efficiency can be increased by
enhancing the gain effect of the boundary sites or the finite Hermitian sites. The main results in this paper, including the implementation scheme of the non-Hermitian lattice with modulated
complex hopping rates, open up a new path for controllable photon transport and provide a promising platform for QIP.

\section*{ACKNOWLEDGMENTS}

Thanks Dr. Yi-Mou Liu for helpful discussions and Dr. Chu-Hui Fan for constructive suggestions. This work is supported by National Natural Science Foundation of China (Grants No. 61378094,
No. 10534002, No. 11674049, and 11704064).

\end{document}